# Integration of a new Cryogenic Liquefier into the IB-1 Cryogenic Test Facility

**Maria Barba[1], Benjamin Hansen[1], Michael White[1], Gregory Johnson[1], Omar Al Atassi[1], Jun Dong[1], Shreya Ranpariya[1], William Soyars[1], Ahmed Faraj[1], Pratik Patel[1], Noelle Besse[2], Annelise Machefel[2] and Lois Perrot[2]**

[1]Fermi National Accelerator Laboratory, Batavia, Illinois 60510, USA

[2]Air Liquide Advanced Technologies, 2 Rue de Cl´emenci`ere, 38360 Sassenage, France

E-mail: mariab@fnal.gov

**Abstract.** The increase over the last years of the testing activities related to quantum systems, SRF cavities for the PIP-II and the LCLS-II projects, as well as superconducting magnets for the HL-LHC project and Fusion research activities, has required the addition of a new Helium cryogenic plant into the existing IB-1 Industrial Cryogenic Test Facility. The new cryogenic plant is composed of a cryogenic liquefier (Cold Box) able to provide up to 340 L/h, a 4 kL Dewar and two Mycom@ compressors providing up to 120 g/s. AL-AT (Air Liquide Advanced Technologies) has taken part of this project by designing and manufacturing the cryogenic liquefier. This new cryogenic plant is connected through a cryogenic distribution system to a 10 kL Dewar, which is part of the existing cryogenic test facility, itself composed of another Cold Box and a Sullair@ compressor. The new cryogenic plant has two main operating modes: one allows to transfer liquid helium at 1.7 bar between the two Dewars, the other allows to transfer supercritical Helium at 2 bar or more between the new Cold Box and the 10 kL Dewar. The entire industrial cryogenic facility is handled by a common Inventory Control System, composed of three control valves, and 9 tanks giving a total buffer volume of more than 1000 m3. This paper presents the technical features of the new Helium cryogenic plant, as well as the main results of the liquefier commissioning phase and details of the helium transfer between the two Dewars, making the connection between the cryogenic plants at the IB-1 Industrial Cryogenic Test Facility.

## 1. Introduction
The Industrial Building 1 (IB-1) is one of the cryogenic industrial test facilities operating at Fermilab. Since 2018, IB-1 has experience an upgrade of its cryogenic facility [1] to support the increasing demand of the cryogenic testing activities for SRF [2, 3, 4] cavities and superconducting magnets [5, 6]. For this upgrade, several components have been incorporated into the existing cryogenic infrastructure in order to be able to increase the helium liquefaction capacity of the entire facility.

## 2. Description of the new cryogenic plant
The IB-1 cryogenic upgrade consists of a new commercial cryogenic helium liquefier (or cold box), a 4 kL liquid helium storage Dewar, a cryogenic distribution system (CDS) connecting the 10 kL Dewar to this 4 kL Dewar and the cold box, two Mycom compressors, 4 buffer tanks and a new gas management system. In Fig. 1 are indicated all the mentioned components (see red square) as well as their connection to the rest of the cryogenic infrastructure already operating at IB-1.

The commercial cryogenic liquefier is an HELIAL cold box, which has been designed and commissioned by AL-AT (Air Liquide Advanced Technologies, Sassenage, France) in October 2022, and is able to provide more than 340 L/h. The liquid helium produced by the cold box is stored in a 4 kL Dewar which is connected to a 10 kL Dewar through the CDS, as shown in Fig. 1. To operate, two Mycom@ compressor skids, each one with its own oil removal system, provide up to 120 g/s at 15 bar(a) to the cold box. All the helium inventory can be stored at room temperature in a tank farm composed of 10 buffer tanks. Finally, the helium circuit is controlled by a gas helium management system which connects the compressors station to the tank farm.

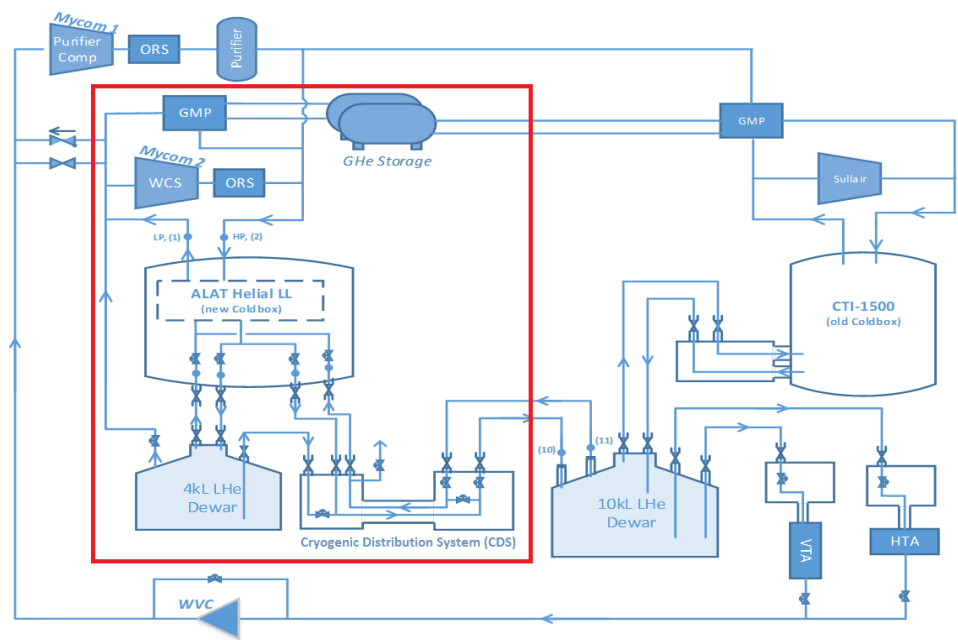

**Figure 1.** Simplified flow diagram.

### 2.1. The HELIAL cryogenic cold box
ALAT HELIAL liquefiers can provide helium liquefaction rate from 35 L/h to 350 L/h @4.5 K. The selected cryo-plant for Fermilab IB1 upgrade is an HELIAL LL (Large Liquefier), it is illustrated in Fig. 2, and its main components are listed below:

- Brazed aluminum heat exchanger cores.
- Cold piping, thermally insulated and located in the coldbox shell that is dynamically pumped by an external primary and secondary vacuum pumping group.
- Liquid nitrogen (LN2) precooling circuit.
- One air regenerable adsorber and one by-pass valve.
- One non regenerable H2/Ne adsorber.
- Two cryogenic expansion turbines on static gas bearings in series.
- One Joule-Thomson valve supplying LHe the 4kL Dewar.

- One Joule-Thomson valve supplying SHe to the CDS.
- Interfaces equipped with bayonets for LHe supply to dewar, SHe supply to CDS, GHe return from dewar and CDS.
- Hard safeties such as safety valves.
- Electrical cabinet.

ALAT also supplied the associated control system and HMI allowing automatic operations of the cold box, including :

- Helium conditioning of the system before cold operations.
- Automatic cooldown and warm-up sequence.
- Control of the liquefaction rate based on, for example, liquid level in the Dewar, buffers pressure, or one external accessible parameter ("CB attenuation coefficient").
- Air adsorber automatic regeneration.
- Software safeties relying on sensors measurements, that are in particular intended to avoid the opening of the hard safeties.
- Digital communication with out-of-scope components (compressors, buffers, CDS).

The HELIAL simplified process flow diagram is illustrated in Fig. 3.

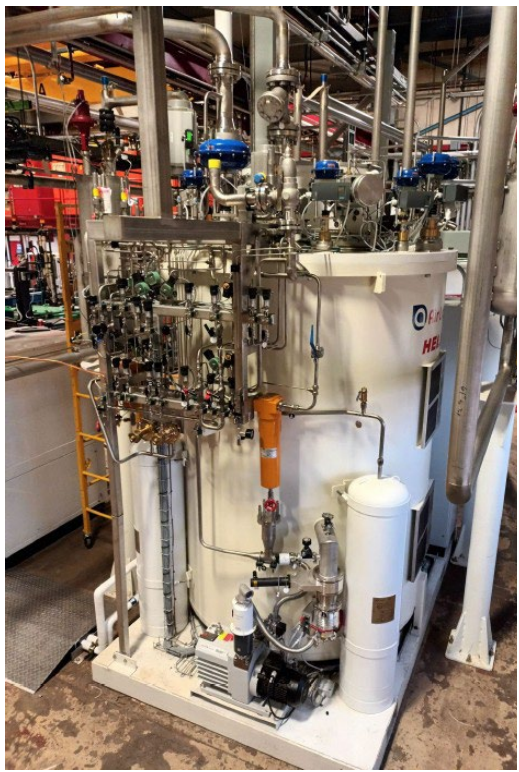

**Figure 2.** Picture of HELIAL Liquefier on the Fermilab site.

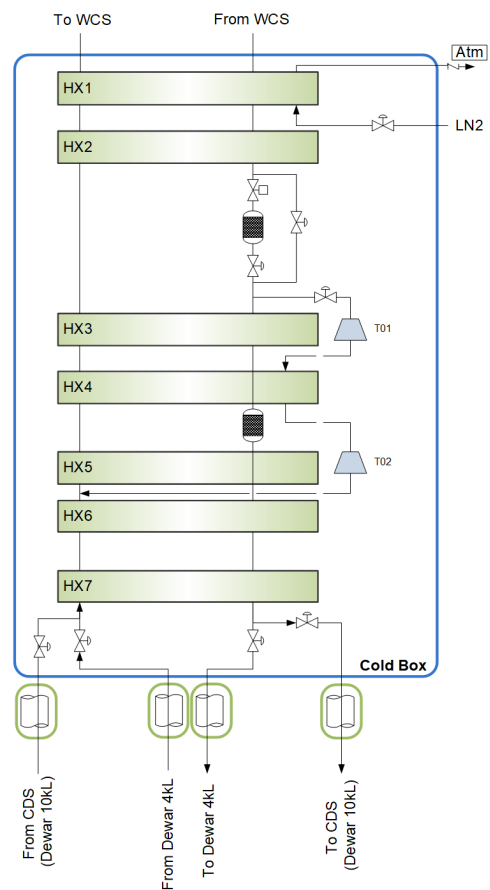

**Figure 3.** HELIAL simplified process flow diagram.

*2.2. The cryogenic distribution system, the tank farm and the helium management system*

The 4 kL Dewar as well as the compressors station have been installed and commissioned during the phase-I of this cryogenic upgrade, and are detailed in [1]. During the second phase, the HELIAL cold box has been commissioned, the total volume of the tank farm has been increased, a new gas management system has been implemented and the cryogenic distribution system (CDS) has been commissioned.

The helium storage capacity of the tank farm, which was initially composed of 6 gas helium buffer tanks (5 storage tanks and 1 quench tank for magnet testing activities), has been increased by adding 4 new buffer tanks. Each tank has a total volume of 113 $m^3$, providing total buffer volume of 1017 $m^3$ that can be increased to 1130 $m^3$ if the quench tank is switched into a storage tank.

Three control valves compose the new gas management system (also known as Inventory Control System), which connects the compressors station to the tank farm and to the cold box, keeping stable high and low pressure helium flows during normal operating conditions and during transient modes such as cool-down or warm-up. The high pressure valve (also called kickback) regulates the high pressure (discharge) entering the cold box by opening at 15.1 bar(a) and releasing any helium excess back into the tank farm. The two low pressure valves (also called make-up and re-circulation) regulate the low pressure (suction) entering the compressors by opening at 1.1 bar(a) and 1.05 bar(a) respectively, and taking more gas helium, if needed, form the tank farm. A simplified diagram of the gas management system is illustrated in Fig. 4, where the control valves are indicated in blue.

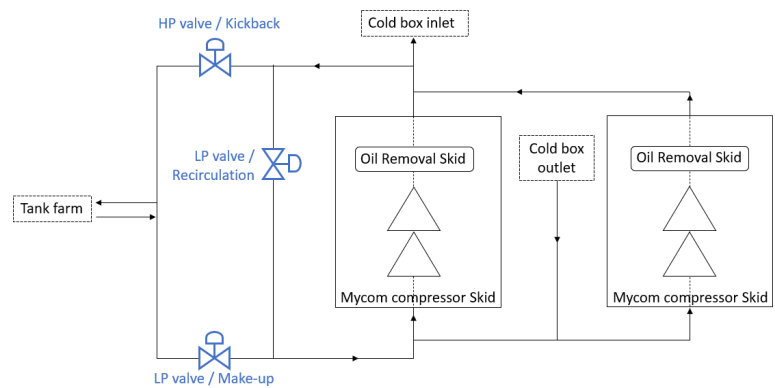

**Figure 4.** Simplified diagram of the gas management system.

The cryogenic distribution system (CDS), makes the connection between the two liquid helium storage Dewars completing the integration of the new components into the existing infrastructure, as illustrated in Fig 5. The CDS is composed of two helium transfer lines and a liquid nitrogen shield, it serves the purpose of transporting cold helium, liquid and vapor, between the Dewars.  It has a total length of 25 m, with bayonet cans at each end (EC-1 and EC-2) and includes and expansion can near the middle for compensation of thermal contraction. It has been designed with a 4 W heat load to all helium circuits and it is equipped with various control valves and instrumentation for operations. It has been fabricated by Ability Engineering Technology (AET) per Fermilab specification, as detailed in [1] and installed during the first phase of the IB-1 cryogenic upgrade but commissioned during the second phase, after the HELIAL cold box.

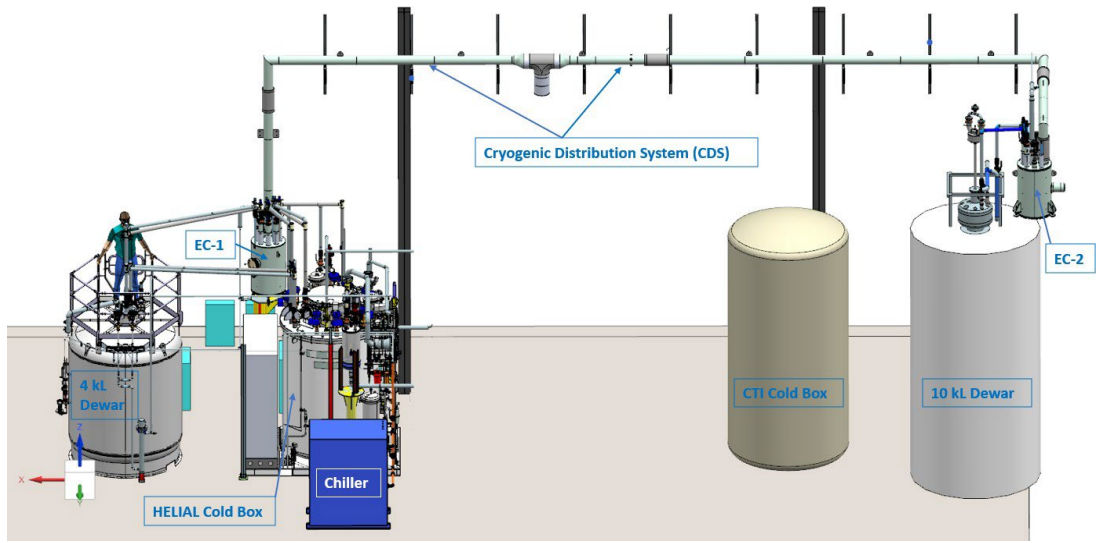

**Figure 5.** Illustration of the cryogenic distribution system connecting both cryogenic plants.

## 3. Performance and integration of the cryogenic plant into the IB-1 Test Facility

This part describes the performance of the cryogenic cold box and the maximum liquid transfer capacity of the cryogenic distribution system between liquid helium storage Dewars.

*3.1. Performance results of the liquefier during the commissioning phase*

During the commissioning phase, the performances of the HELIAL were proved during a 3 h rising level test in the 4 kL Dewar. Already cold (Dewar half-full), the cold box was left producing liquid helium at its maximal capacity. Pressure in the Dewar was regulated with a PID controller on the Dewar outlet valve, releasing gas to the cold line of the cold box. The measurement of the liquid level was made with a superconducting liquid helium probe supplied by Fermilab.

The performance test results obtained during the commissioning of the HELIAL cold box are indicated in table 1.

**Table 1.** Performance test results during commissioning of the HELIAL cold box.

| Tested criteria | Liquefaction capacity (L/h) |
|---|---|
| Date | 11/10/2022 |
| Performance Guaranteed (based on Fermilab requirements) | 250 L/h |
| Expected performance according to AL-AT calculation | 350 L/h |
| Initial test conditions | 4 kL liquid level = 51.82 % (2072.8 L) |
| Final test conditions | 4 kL liquid level = 77.77 % (3110.8 L) |
| Test duration | 3 hours |
| Result | 346 L/h |

*3.2. Liquid helium transfer capacity between cryogenic plants*

For the commissioning of the cryogenic distribution system, a liquid helium transfer capacity test has been performed, transferring helium from the 4 kL Dewar to the 10 kL Dewar while both cold boxes where in liquefaction mode and magnet testing activities [6] were ongoing at the IB-1 facility.

The test was divided in 3 phases: the helium transfer with attenuation of the HELIAL cold box phase, the very limited helium transfer phase and the maximum helium transfer phase. Fig 7 shows the evolution of the volume of liquid helium in both Dewars during the entire test. At the beginning of the test, the 4 kL Dewar was full with 1652 L and the 10 kL Dewar was full with 4637 L. The first phase of the test focused on the cooldown of the CDS, and the HELIAL cold box was attenuated to 80 % on average to don't overfill the Dewar too quickly. The cooldown process of the CDS took place during the first minutes of this phase and is described as follows: first, valves PCV-3414, PCV-3422 and PCV-3417 (indicated in Fig. 6) open to 100 % allowing a helium path from the 4 kL Dewar through the CDS and returning to the compressor station, and cooling down most of the CDS system. Secondly, PCV-3416 and PCV-3415 are open up to 40 % to use the boils-off of the 10 kL Dewar to cooldown the two lines between the CDS and the 10 kL Dewar while reducing PCV-3414 to 10 %. Then, when the CDS is cold, the helium transfer between Dewars can start by opening to it's maximum PCV-3414, PCV-3415 and closing PCV-3422. PCV-3416 will adapt to keep a stable pressure in the 10 kL Dewar. Once the CDS is cold, after a few minutes of the beginning of the first phase, it can be noticed on Fig. 7 that the volume of liquid helium in the 10 kL Dewar starts to rise due to the transfer and despite the ongoing testing activities. During the second phase it has been decided to limit the helium transfer in order to build up more liquid helium inventory in the 4 kL Dewar while keeping the CDS cold (PCV-3414 only open at 10 %) because the volume of testing activities decreased. This led to a stabilization of the volume of liquid helium in the 10 kL Dewar during this second phase. During the third phase, the helium needs to support testing activities increased again, then both cold boxes were producing liquid at 100 % of their capacity and the helium transfer was maximum. At the beginning of this third phase, the volumes of liquid helium in the 4 kL and 10 kL Dewars were 1745 and 5385 L, respectively. At the end of this phase, they were 1086 and 4944 L, respectively. The HELIAL cold box was at its maximum capacity during this entire third phase (producing 346 L/h based on the HELIAL commissioning results), the difference of the volume of liquid helium in the 4 kL Dewar between the beginning and the end of this phase was 659 L and the phase last 2.5 hours. This makes an average liquid helium transfer per hour of more than 609 L, confirming the high efficiency of the helium transfer process between Dewars through the CDS.

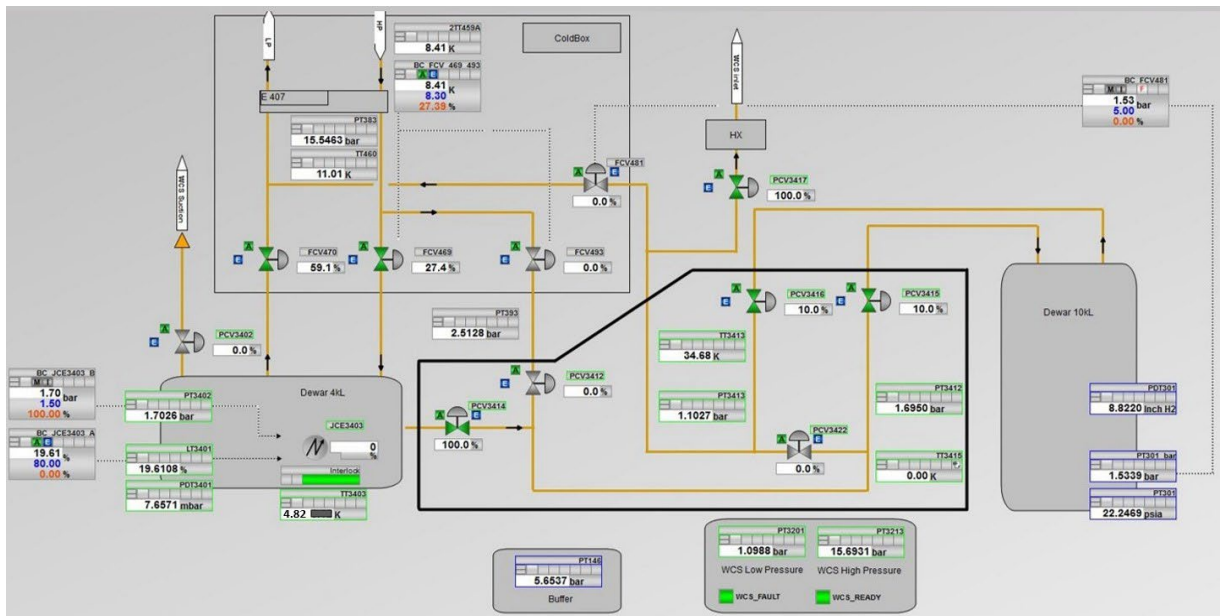

**Figure 6.** Schematic illustration of the CDS including both Dewars and the instrumentation of the system.

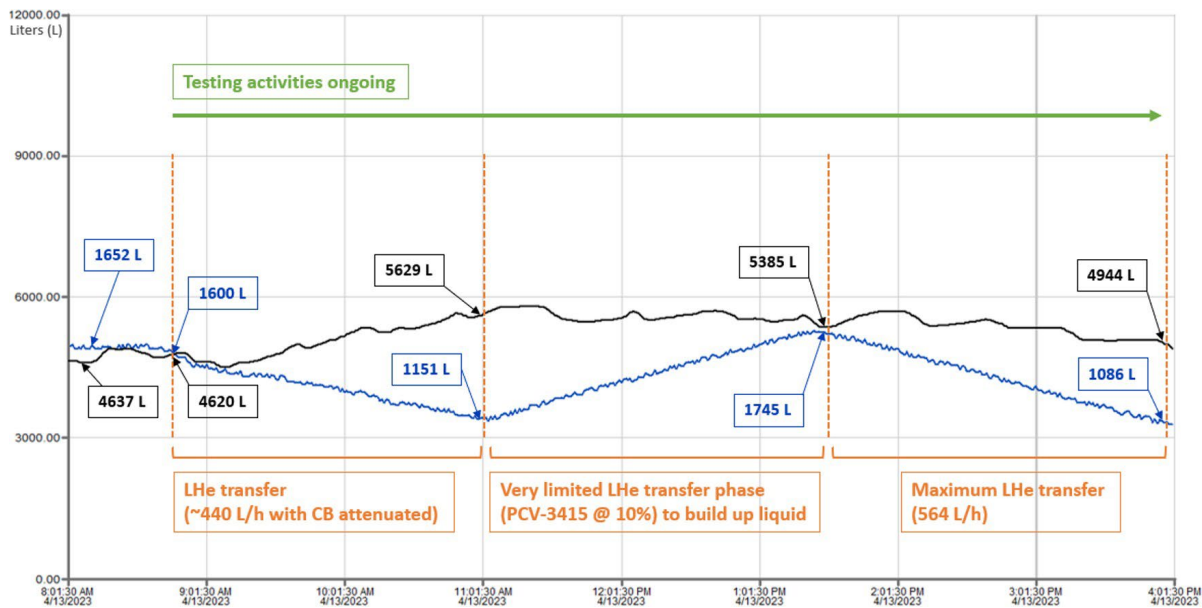

**Figure 7.** Evolution of the liquid helium level in the Dewars during LHe transfer phases from the 4 kL Dewar (-) to the 10 kL Dewar (-).

## 4. Conclusion

The second phase of the IB-1 cryogenic upgrade has been completed with the addition of buffer tanks to the tank farm, the integration and commissioning of a new gas management system, and the commissioning of the HELIAL cryogenic cold box and the cryogenic distribution system. The different components of this cryogenic upgrade project have been integrated into the existing IB-1 infrastructure, increasing the helium liquefaction capacity in more than 340 L/h and allowing a liquid helium transfer between the two cryogenic plants of more than 600 L/h. This upgrade provides the extra capacity and system reliability needed to support Fermilab's aggressive project goals.

## Acknowledgments
The authors of this paper would like to acknowledge all the cryogenic operators of the IB-1 Test Facility: George W. Kirschbaum, Daniel B. Marks, Randall S. Ward, Brett T. Swanson and Ernest Feret, as well as the rest of the IB-1 technical team for their availability and dedication to this project.

This manuscript has been authored by Fermi Research Alliance, LLC under Contract No. DE-AC02-07CH11359 with the U.S. Department of Energy, Office of Science, Office of High Energy Physics.